\documentclass[conference]{IEEEtran}
\IEEEoverridecommandlockouts
\usepackage{tabularx}
\usepackage{soul,color}
\usepackage{cite}
\usepackage{amsmath,amssymb,amsfonts}
\usepackage{algorithmic}
\usepackage{graphicx}
\usepackage{booktabs}
\usepackage{multirow}
\usepackage{tabularx}
\usepackage{caption}
\usepackage{textcomp}
\usepackage[aboveskip=5pt]{caption}
\usepackage[belowskip=-8pt]{caption}
\usepackage{soul}
\usepackage{graphicx}
\usepackage{bm}
\usepackage{mathrsfs}
\usepackage{svg}
\graphicspath{{figs/}{./}}

\AtEndDocument{\nocite{*}}

\def\BibTeX{{\rm B\kern-.05em{\sc i\kern-.025em b}\kern-.08em
    T\kern-.1667em\lower.7ex\hbox{E}\kern-.125emX}}
\begin{document}
\title{

QLook: Quantum-Driven Viewport Prediction for Virtual Reality\\

}
\author{Niusha Sabri Kadijani,  
Yoga Suhas Kuruba Manjunath,
Xiaodan Bi,
Lian Zhao,\\
Department of Electrical, Computer \& Biomedical Engineering,\\ Toronto Metropolitan University, Toronto, Canada.\\
\
Email: \{niusha.sabrikadijani@torontomu.ca, yoga.kuruba@torontomu.ca, \\xiaodan.bi@torontomu.ca l5zhao@torontomu.ca\}
}

\maketitle

\begin{abstract}
We propose QLook, a quantum-driven predictive framework to improve viewport prediction accuracy in immersive virtual reality (VR) environments. The framework utilizes quantum neural networks (QNNs) to model the user movement data, which has multiple interdependent dimensions and is collected in six-degree-of-freedom (6DoF) VR settings. QNN leverages superposition and entanglement to encode and process complex correlations among high-dimensional user positional data. The proposed solution features a cascaded hybrid architecture that integrates classical neural networks with variational quantum circuits (VQCs)-enhanced quantum long short-term memory (QLSTM) networks. We utilize identity block initialization to mitigate training challenges commonly associated with VQCs, particularly those encountered as barren plateaus. Empirical evaluation of QLook demonstrates a 37.4\% reduction in mean squared error (MSE) compared to state-of-the-art (SoTA), showcasing superior viewport prediction.
\end{abstract}

\vspace{-1em}

\section{ Introduction}

The demand for immersive technology continues to grow due to its applications in various fields~\cite{zhang2022realvr}. Delivering high-quality virtual reality (VR) experiences, which are realized by head-mounted displays (HMDs) and handheld controllers, requires overcoming significant technical challenges, including achieving adaptive data rates, ultra-low latency, and dynamic system configurations. Seamless immersion experience for users is vital in VR to avoid cybersickness, which occurs when the latency of the video frame rendering exceeds 20 ms, corresponding to the user’s physical action \cite{yao2014oculus}. Ultra-low latency is one of the critical factors achieved by buffering the required video frame ahead of time. Viewport, the visible portion of the video frame to the user on an HMD screen, and its prediction helps buffer the required video frame. Predicting the viewport to render future frames to avoid cybersickness is essential, especially in six degrees of freedom (6DoF) environments.  In such settings, both the position and orientation of the user in 3D space generate multiple interdependent data streams.

An object detection and trajectory-based approach for 360\textdegree{} viewport prediction is proposed in \cite{jin2023ebublio}. The passive-aggressive regression model relies on the assumption that users with similar past trajectories share the same viewport. However, this assumption limits its generalization capability and adaptability to unseen user behavior. 
Huang \textit{et al.} introduce a contextual multi-armed bandit model to predict viewport \cite{huang2023predictive}. The model relies on empirically observed reward signals. These signals are context-specific and cannot be estimated in advance for new users or unfamiliar scenarios, which limits the model's applicability. Li \textit{et al.} consider tile-based viewport prediction using long short-term Memory (LSTM) \cite{li2023user}. However, sequential models often struggle to capture complex temporal dependencies in short sequences, requiring access to more data for practical applicability. A deep reinforcement learning-based solution is proposed in \cite{bi2024two} to predict the viewport by leveraging the frame grouping method. However, its inference time may not satisfy the stringent ultra-low latency requirements of VR applications. In another study, a viewport prediction solution is proposed using head and body pose \cite{hou2020motion}. Nevertheless, positional and head orientation data are treated individually, which neglects the interdependence between them and oversimplifies the problem to be handled by classical neural networks, leading to higher prediction error and reduced accuracy.
Ouellette \textit{et al.} present the only existing GRU-based model for 6DoF viewport prediction using head and body trajectories \cite{ouellette2025ll}. We consider it the state-of-the-art (SoTA) for fine-grained 6DoF prediction using temporal modeling, as it is the most recent method available, despite its limitations in generalization and accuracy.

Most of the existing works focus on 360\textdegree{} video data, where users explore the environment only through head rotation. Prior studies ignore the 6DoF VR scenario, where users move freely in both position and orientation. Conventional approaches treat positional and head orientation data independently. Modeling them separately fails to capture the mutual influence between body and head movements in 6DoF VR. Sequential models such as LSTM require long sequences and large parameter sets to learn complex temporal dependencies. In 6DoF VR, where sequences are short and data is noisy, prediction errors propagate through time, degrading performance.

Quantum neural networks (QNNs) powered by superposition and entanglement offer promising capabilities for modeling high-dimensional and interdependent user movement data in 6DoF VR viewport prediction. Superposition enables quantum parallelism to explore complex feature spaces efficiently. Entanglement introduces strong correlations between quantum states, allowing the model to capture non-local dependencies inherent in user movement patterns, where body position and head orientation influence each other dynamically. Quantum long short-term memory (QLSTM) is a well-known QNN that performs well in identifying patterns in complex and high-dimensional temporal data \cite{emu2024quantum}. The QLSTM architecture introduced by PennyLane is implemented using Variational quantum circuits (VQCs) \cite{bergholm2020pennylane}. VQCs provide a robust error correction of classical optimizers in noisy intermediate-scale quantum (NISQ) devices \cite{cerezo2021variational}. However, VQCs struggle with barren plateaus problems, especially during training \cite{cerezo2021cost}. Therefore, we study the QLSTM for viewport prediction in a complex and high-dimensional 6DoF VR setup. 

To address the research gaps, we propose a QLook framework with a cascade quantum-driven approach for viewport prediction in 6DoF VR by combining body positioning data with head orientation. The first stage of the proposed cascade model is formed using an LSTM, which predicts the positional data. The predicted positional data is then combined with head orientation data to form an input to the second stage of the cascade model formed using a QLSTM. We conduct a comprehensive empirical evaluation of our approach against two baselines, including a cascade classical LSTM model, using a real-world dataset \cite{baldoni2024questset}. Our results demonstrate a 37.4\% improvement in viewport prediction accuracy compared to the SoTA \cite{ouellette2025ll}.
To the best of our knowledge, our's is the first work to study QLSTM for viewport prediction in 6DoF VR setting.
\vspace{-1em}
\section{System Model}
\begin{figure}
    \centering
     \includegraphics[width=\linewidth]{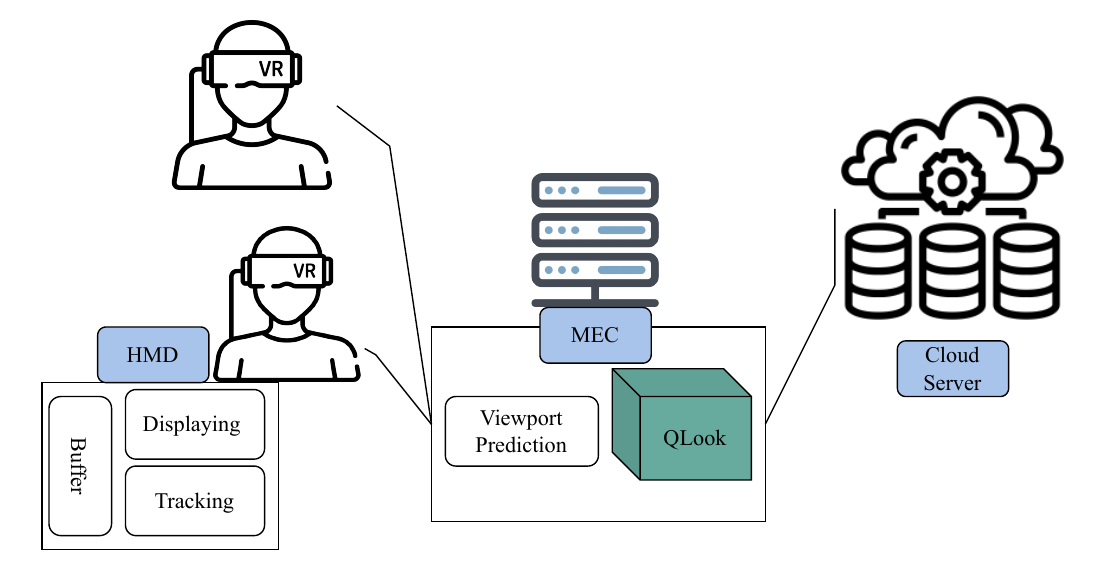}
    \caption{System model}
    \label{fig:system_model}
    \vspace{-1em}
\end{figure}

\vspace{-0.3em}
The system model of the solution is shown in Fig. \ref{fig:system_model}. The Multi-access edge computing (MEC) serves multiple VR users. The HMD displays the view portion of the virtual environment while continuously tracking the user's gaze. The controllers provide additional positional data. HMD transmits the captured sensor data to the MEC, where the QLook framework predicts the viewports. If the frames are already cached at MEC, once the viewport is predicted, the MEC renders the corresponding frame and transmits the frames to HMD. The accurate viewport prediction enables corresponding frames to be pre-buffered and rendered to the HMD to avoid cybersickness and provide a seamless experience.

\vspace{-0.5em}
\subsection{Problem Formulation}
\vspace{-0.1em}
In a 6DoF VR setting, predicting a user's future viewport involves estimating both spatial position and rotational orientation over time. In each time step \( t \), a data point is defined as a 32-dimensional vector given as,
\vspace{-0.4em}
\begin{equation}
\mathbf{m}_t = (\tau_t, \mathbf{p}_t, \mathbf{e}_t, \mathbf{c}_t, \mathbf{a}_t),
\end{equation}

\noindent where \( \tau_t \) is the timestamp; \( \mathbf{p}_t \in \mathbb{R}^3 \) represents the 3D position in coordinates \( (x, y, z) \); \( \mathbf{e}_t \in \mathbb{R}^3 \) denotes the head orientation in Euler angles (pitch, roll, yaw); \( \mathbf{c}_t \in \mathbb{R}^3 \) represents the controller position; and \( \mathbf{a}_t \in \mathbb{R}^{23} \) includes other features, such as touch button states and derived features, to enrich the dataset representation.

For the first prediction stage, we define an input vector:
\vspace{-0.5em}
\begin{equation}
\vspace{-0.5em}
\mathbf{q}_t = (\tau_t, \mathbf{p}_t),
\vspace{-0.1em}
\end{equation}
\noindent and the corresponding input matrix formed by the past \( n \) time steps from \( t_k \) to \( t_{k-n+1} \), given as:
\vspace{-0.8em}
\begin{equation}
\vspace{-0.5em}
\mathbf{Q}_{t_k} = [\mathbf{q}_{t_k}, \mathbf{q}_{t_{k-1}}, \dots, \mathbf{q}_{t_{k-n+1}}],
\vspace{-0.2em}
\end{equation}
\noindent where $k$ denotes the starting index of the input sequence window.
The first predictive function \( \psi \) estimates the future user position in three axes $(x,y,z)$:
\vspace{-0.5 em}
\begin{equation}
\vspace{-0.5 em}
\label{eq:psi}
 \hat{\mathbf{p}}_k = \psi(\mathbf{Q}_{t_k}).
\end{equation}

After predicting \( \hat{\mathbf{p}}_k \), the predicted position is concatenated with other features, and state as the input vector:
\vspace{-0.5 em}
\begin{equation}
\vspace{-0.5 em}
\mathbf{m}_{t_{k+n}} = (\hat{\mathbf{p}}_{t_{k+n}}, \mathbf{e}_{t_{k+n-1}}, \mathbf{c}_{t_{k+n-1}}, \mathbf{a}_{t_{k+n-1}}),
\end{equation}

\noindent and we define the full input matrix for the second prediction stage as:
\vspace{-1em}
\begin{equation}
\vspace{-0.5em}
\mathbf{M}_{t_k} = [\mathbf{m}_{t_k}, \mathbf{m}_{t_{k-1}}, \dots, \mathbf{m}_{t_{k-n+1}}]\textcolor{red}{.}
\end{equation}

The second predictive function \( \phi \) uses this sequence to predict the future head orientation represented by Euler angles in the axes of $(x,y,z)$:
\vspace{-1em}
\begin{equation}
\label{eq:phi}
    \hat{\mathbf{y}}_k =\phi(\mathbf{M}_{t_k})\textcolor{red}{.}
\vspace{-0.5em}
\end{equation} 

The proposed framework solves the Eq. (\ref{eq:psi}) and (\ref{eq:phi}) to find the predictive functions.

\section{Preliminaries of Quantum LSTM}
\label{sec:quantum} 

\subsection{Variational Quantum Circuit}
\begin{figure}
    \centering
    \label{fig: VQC}
     \includegraphics[width=\linewidth]{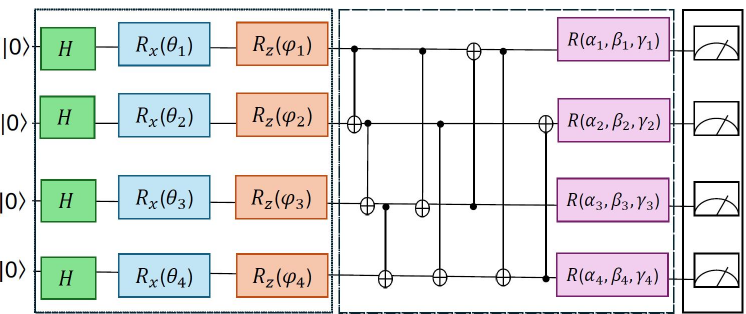}
    \caption{The architecture of the VQC, where $(\theta_i)$ and $(\phi_i)$ are encoded from the classical input data, while $(\alpha_i, \beta_i, \gamma_i)$ for $i=1,2,3,4$ are trainable parameters,follows the design introduced in \cite{chen2022quantum}.}
    \label{fig:VQC}
    \vspace{-0.5cm}
\end{figure}

Superposition and entanglement are the distinguishable properties of quantum computing. Quantum computing uses qubits. A qubit can have one state or a combination of basis states. The linear combination of basis states as a quantum state is called a superposition \cite{bouwmeester2000physics}. The correlation among the quantum states remains intact, regardless of how far the qubits are entangled \cite{bouwmeester2000physics}. 
NISQs are the current generation of quantum computers. The absence of gate error correction and limitations of the number of qubits are the drawbacks of the NISQ~\cite{cerezo2021cost} that limit the applicability of NISQ to large-scale quantum computations. VQCs provide a practical approach for leveraging NISQ devices for large-scale quantum computations, including quantum eigensolvers~\cite{grant2019initialization} and quantum neural networks. VQCs, as quantum circuits, are hardware-efficient. The shallow depth of the parameterized quantum circuits mitigates the lack of error correction of NISQs \cite{cerezo2021variational}. VQCs delegate the training process of the parameterized part to a classical gradient-based optimizer, mitigating the limitations of NISQ devices. VQCs perform more efficiently with short gate sequences, making them particularly suitable for NISQ devices~\cite{schuld2019quantum}.
A typical VQC is illustrated in Fig. \ref{fig:VQC}. The VQC consists of three main components. The first component is the encoding layer, which maps classical input data to a quantum state using angle encoding, which is implemented using Hadamard, Pauli-X, and Pauli-Z gates. The second component is the parameterized quantum layer, which contains trainable quantum gates with tunable parameters that are optimized during training using classical gradient-based methods. The final component is the measurement layer, which collapses the quantum states to produce classical outputs. These outputs, expectation values, are then used by classical models that are integrated into hybrid quantum-classical models. 

\subsection{QLSTM Cell}
\vspace{-0.3em}

By integrating VQC into the internal memory dynamics of the gating mechanism of the LSTM \cite{chen2022quantum}, QLSTM enables the encoding of information in Hilbert space, which offers expressive capacity and potential performance advantages.
In a classic LSTM, fully connected layers typically handle each linear transformation within the gating functions. In contrast, QLSTM, they are replaced with VQC modules, each acting on the concatenated input vector $v_t = [h_{t-1}, x_t]$
where $h_{t-1}$ is the hidden state from the previous time step, and $x_t$ is the current input.
The QLSTM cell performs the following operations at each time step $t$:
\vspace{-0.5 em}
\begin{align}
{f_t} &= {\sigma(\text{VQC}_1(v_t))}, &&{\text{forget gate}} \\
{i_t} &= {\sigma(\text{VQC}_2(v_t))}, &&{\text{input gate}} \\
{\tilde{C}_t} &= {\tanh(\text{VQC}_3(v_t))}, &&{\text{cell candidate}} \\
{c_t} &= {f_t \odot c_{t-1} + i_t \odot \tilde{C}_t}, &&{\text{cell state update}} \\
{o_t} &= {\sigma(\text{VQC}_4(v_t))}, &&{\text{output gate}} \\
{h_t} &= {\text{VQC}_5(o_t \odot \tanh(c_t))}, &&{\text{hidden state}} \\
{y_t} &= {\text{VQC}_6(o_t \odot \tanh(c_t))}, &&{\text{output}}
\vspace{-0.5 cm}
\end{align}
\vspace{-0.1 em}
\noindent where, $\sigma$ and $\tanh$ denote the sigmoid and hyperbolic tangent activation functions, respectively. Each VQC block processes its input through a parameterized quantum circuit, followed by quantum measurement, resulting in a vector of expectation values. The measured outputs serve as quantum counterparts to the classical weight matrices and biases in LSTM, providing a quantum-enhanced nonlinear transformation before being passed through classical activations.

\section{Methodology}
\vspace{-0.5em}
\subsection{Model Architecture}
\begin{figure}
    \centering
     \includegraphics[width=\linewidth]{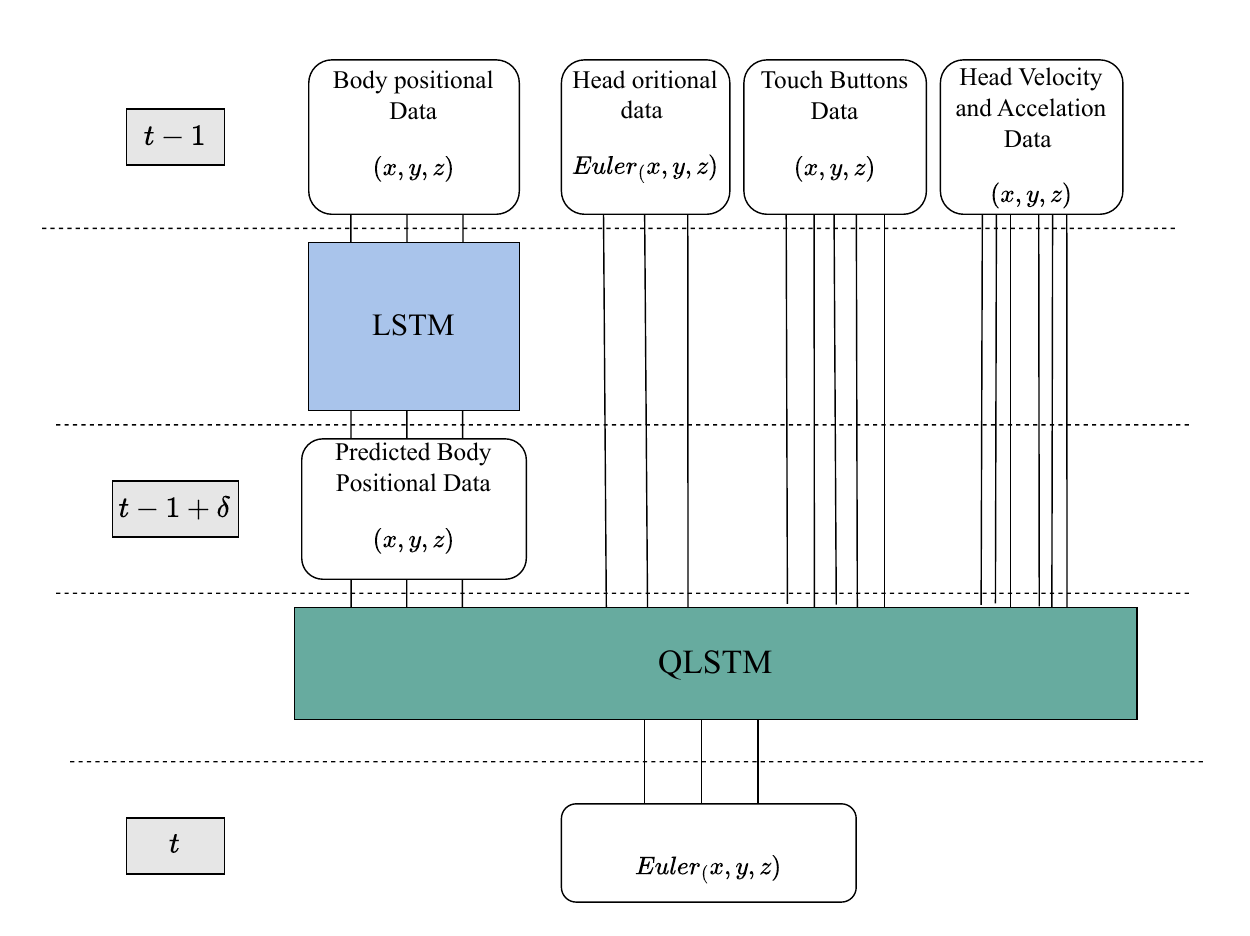}
    \caption{Demonstration of QLook architecture}
    \label{fig:cascade_model}
    \vspace{-1em}
\end{figure}

We propose QLook, a cascade model built on quantum-driven deep neural networks to solve the Eq. (\ref{eq:psi}) and (\ref{eq:phi}). As shown in Fig. \ref{fig:cascade_model}, the QLook architecture is built on two modules. The first module is a stacked standard LSTM network, selected for its faster inference time. The first module is trained to predict positional data along the $(x,y,z)$ axes to solve the Eq. (\ref{eq:psi}). The QLSTM module, as the second module, solves Eq. (\ref{eq:phi}).  The predicted body positional data from the first module, along with controller data and two more extracted features: head velocity and head acceleration, are an input to the QLSTM module.  The output of the cascade model provides orientation data represented by Euler angles on three axes. 
\vspace{-0.5em}
\subsection{Data Augmentation and Initialization}
 Gaussian noise is integrated into the training data during the preprocessing phase. It is sampled from a zero-mean normal distribution \( \mathcal{N}(0, \sigma^2) \), where the standard deviation \( \sigma \) is empirically set to 2 based on initial experiments. Gaussian noise is a widely adopted approach in the literature for modeling random sensor errors and simulating real-world uncertainties in machine learning applications. This augmentation increases the complexity of the dataset and encourages the QLSTM module to learn more meaningful patterns rather than memorizing specific details. Consequently, it enhances model robustness and improves generalization, enabling better handling of variability in unseen VR data.

QLSTM architecture is built on the VQC, which is prone to poor initialization. VQCs are especially susceptible to the barren plateaus problem, in which gradients vanish as circuit depth increases \cite{grant2019initialization}. To mitigate the impact of inadequate initialization in QLSTM, we employ the identity block initialization technique, originally presented by \cite{grant2019initialization}. The parameterized portion of a standard VQC, as shown in Fig. \ref {fig:VQC}, is usually initialized randomly without enforcing identity preservation, which frequently results in vanishing gradients. In contrast, each variational block is constructed as a pair of mirrored layers. Each layer applies parameterized single-qubit rotations of the form $R(\alpha_i, \beta_i, \gamma_i)$ to every qubit. Layer~1 uses parameters $(\alpha_i, \beta_i, \gamma_i)$, while Layer~2 is initialized with $(-\alpha_i, -\beta_i, -\gamma_i)$, ensuring that the combined operation satisfies $R(\alpha_i, \beta_i, \gamma_i) \cdot R(-\alpha_i, -\beta_i, -\gamma_i) = I$.

 \noindent The quantum layer becomes easily trainable by initializing this part as an identity operation, which enhances the model’s overall convergence. 

\vspace{-0.5em}
\section{Results and Discussion}
\vspace{-0.5em}
\subsection{Model Design and Implementation Details}

Meta Quest 2 is used as HMD with 90 frames per second, which allows us to predict the viewport in a window of 11.11 milliseconds. Given this constraint, the model should predict the viewport accurately within the time requirements. Therefore, we employ the LSTM module to guarantee the real-time performance. Given the computational demands on graphical processing units (GPUs), the inference time of QLSTM remains slower than that of LSTM. Therefore, we intentionally design the first module as an LSTM to ensure fast inference and meet the real-time constraints of the device.

We implement the proposed framework using Python 3.9.20 with PyTorch  2.5.1 and PennyLane 0.38 for quantum circuit simulation. All experiments are conducted on a server equipped with dual NVIDIA RTX A5000 GPUs with 24 GB VRAM each and 128 GB RAM. 
The detailed hyperparameters and model configurations are summarized in Table~\ref{tab:hyperparameters}.

\begin{table}[ht]
\centering
\caption{Hyperparameters and architecture details of the LSTM and QLSTM modules}
\label{tab:hyperparameters}
\begin{tabular}{@{}lcc@{}}
\toprule
\textbf{Parameter} & \textbf{Module 1 (LSTM)} & \textbf{Module 2 (QLSTM)} \\
\midrule
Training epochs          & 50                     & 50                         \\
Learning rate            & 0.001                  & 0.001                      \\
Sequence length          & 10                     & 10                         \\
Hidden units             & 32                     & 100                        \\
Number of layers         & 2                      & 2                          \\
Batch size               & 32                     & 64                         \\
Dropout rate             & 0.2                    & 0.2                        \\
Optimizer                & Adam                   & Adam                       \\
Quantum components       & --                     & 4 qubits, 2 VQC layers     \\
\bottomrule
\end{tabular}
\vspace{-1em}
\end{table}
\vspace{-1em}
\subsection{Dataset Details}

We use the Questset dataset~\cite{baldoni2024questset}, which provides 6DoF movement data capturing users’ head orientational and body positional data during VR sessions. 
This dataset is highly suitable for our work as it provides realistic user movement data in immersive VR environments, allowing us to evaluate the robustness and generalization ability of the proposed QLook framework under varying interaction speeds and motion patterns. While multi-user VR scenarios would better reflect real-world conditions, such datasets are currently unavailable.

In total, the part of the dataset that we use for this study contains approximately 477,000 samples. We chose users 2 to 8 from order 1 group 1 and 2, for more detailed information, refer to the dataset \cite{baldoni2024questset}. We use 42\% of the available samples for our experiments. Within this subset, we apply a 60\%-20\%-20\% split for training, validation, and testing, respectively. We segment the data into sequential windows using a sliding window approach, preserving temporal dependencies within each sequence.

\vspace{-0.5 em}
\subsection{Performance Comparison and Quantitative Evaluation}

We implement the cascade classical LSTM with two LSTM modules as a baseline with the same hyperparameters as the QLook framework. We evaluate the prediction performance using the mean squared error (MSE) computed on the test set after training. 
QLook achieves 0.087 MSE. The cascade classical model obtains 0.109 MSE on the test set of the dataset, indicating better predictive performance of the quantum-driven framework.
Simulating quantum circuits on GPUs is computationally intensive; training the QLook framework takes 170,991.85 seconds, while the cascade classical LSTM finishes in just 3,458.16 seconds. As Table \ref{tab:parameters} shows, Qlook has a significantly smaller number of trainable parameters, approximately 99\% fewer than the cascade classical LSTM. This reduction suggests that, when executed on dedicated quantum hardware, the inference time could be drastically reduced, as the quantum supremacy suggests requiring fewer gates and queries than classical models~\cite{biamonte2017quantum}.

\begin{table}[ht]

\caption{{Comparison of QLook and Baseline Models}}
\label{tab:parameters}
\centering
\begin{tabularx}{0.48\textwidth}{lXrrr}
\toprule
\textbf{Model} & \textbf{Description} & \textbf{MSE} & \textbf{Improvement} & \textbf{\# Parameters} \\
\midrule
QLook  & Cascade classical-quantum model & \textbf{0.087} & ---- & \textbf{1,335} \\
\midrule
\multirow{2}{*}{\textbf{Baselines}} 
& Cascade classical model & 0.109 & \textbf{20.11\%} & 134,703 \\
& {SoTA~\cite{ouellette2025ll}} & {0.140} & {37.4\%} & {40,503} \\
\bottomrule
\end{tabularx}
\end{table}
\vspace{-0.5 em}

\begin{table}[ht]
\centering
\caption{{RMSE Comparison of Models}}
\label{tab:rmse_euler}
{
\begin{tabular}{lccc}
\toprule
\textbf{Model} & \textbf{Euler X} & \textbf{Euler Y} & \textbf{Euler Z} \\
\midrule
QLook             & \textbf{0.253} & \textbf{0.028} & \textbf{0.438} \\
Cascade Classical & 0.289 & 0.168 & 0.456 \\
SoTA \cite{ouellette2025ll} & 0.333 & 0.100 & 0.510 \\
\bottomrule
\end{tabular}
}
\end{table}

\begin{figure}[t]
    \centering
     \includegraphics[width=\linewidth]{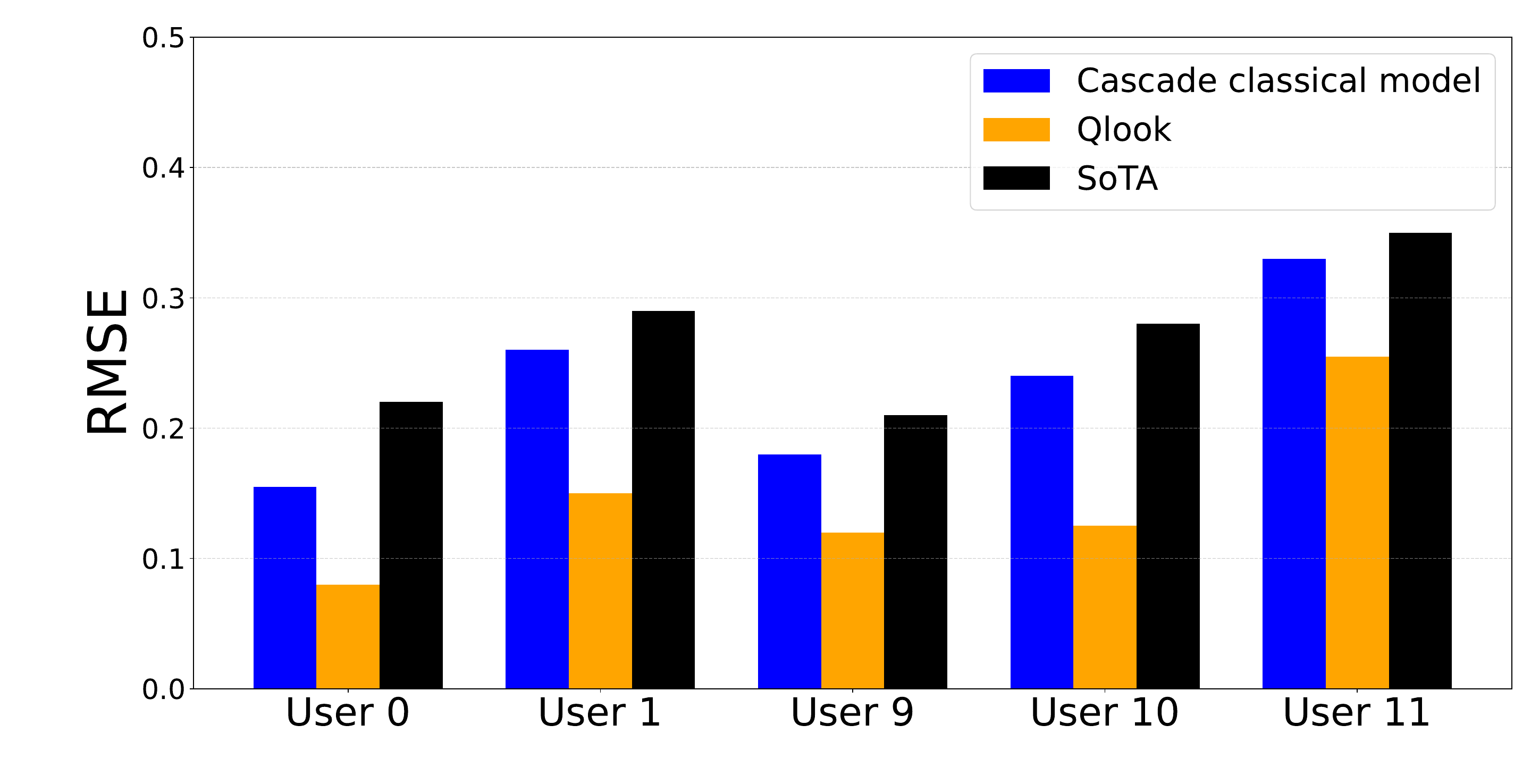}

    \caption{{RMSE comparison across selected users, showing QLook’s lower error rates compared to the baselines.}}
    \label{fig:rmse_quantumenhanced}
\end{figure}

\begin{figure}[t]
    \centering
    \label{euler_z}
     \includegraphics[width=\linewidth]{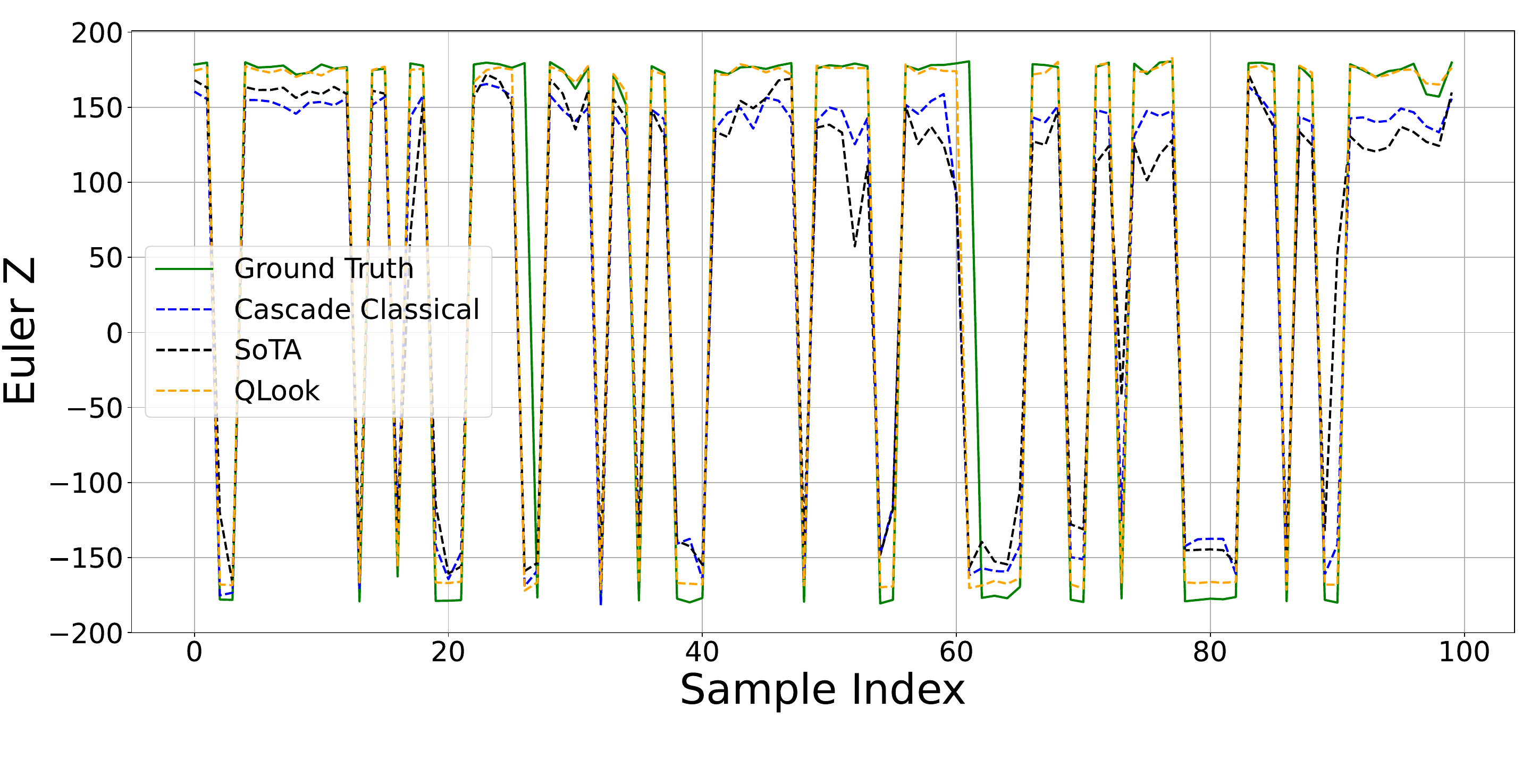}
    \vspace{-1.0 cm}
    \caption{{Comparison of Euler Z predictions over 200 samples. QLook (orange dashed) closely tracks the ground truth (green solid) compared to the baseline.}}
    \label{fig:eulerz}
    \vspace{-0.5 em}
\end{figure}

\vspace{-0.4 cm}
An important aspect of the robustness and generalization capability of the proposed model is demonstrated in Fig. \ref{fig:rmse_quantumenhanced}. We define generalization as the model's ability to accurately predict head and body positional patterns of unseen users' data that are not included in the training phase. To provide a representative evaluation, the trained model is tested on five users: 0, 1, 9-11, selected from order 1, groups 1 and 2, based on the user characteristics outlined in the survey of \cite{baldoni2024questset}. We select the data from these users to capture diverse behavioral patterns relevant to VR scenarios.
\noindent The superior predictive performance of QLook is evident in table \ref{tab:rmse_euler}, highlighting its generalization ability across unseen user data. Fig. \ref{fig:eulerz} shows that QLook more closely follows the pattern of the Euler Z and preserves temporal dynamics more accurately than baselines. 

\subsection{Identity Block Effect}

\begin{figure}[t]
    \centering
     \includegraphics[width=\linewidth]{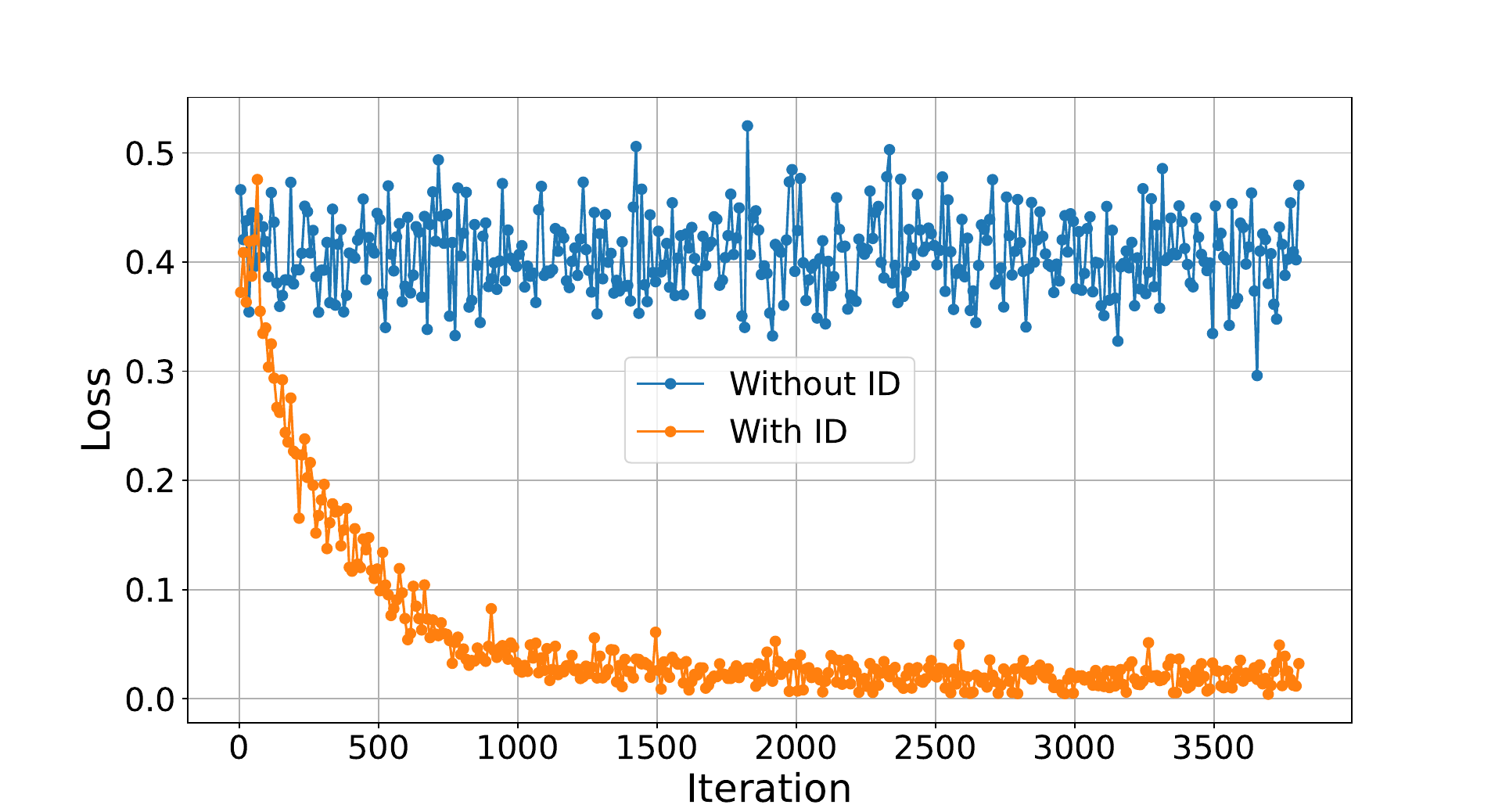}
    \caption{Effect of the identity block on the convergence behavior of the QLSTM module. Adding the identity block helps the model converge faster and reach a lower loss compared to the standard QLSTM without identity initialization.}
    \label{fig:identity_block}
    \vspace{-0.5 cm}
\end{figure}

The QLSTM module within the QLook framework faces the challenge of barren plateaus during training, slowing down the optimization of quantum parameters and convergence. This behavior is observable in Fig. \ref{fig:identity_block}, where the training loss stays at higher values. The identity block is introduced within the QLSTM architecture and acts as an effective solution to this problem, facilitating a smoother convergence and improving the model's learning capability. Fig. \ref{fig:identity_block} shows the iteration-level training loss per mini-batch over one epoch, comparing the QLSTM module performance with and without the identity block. The results demonstrate that incorporating the identity block leads to faster convergence and a more stable training process.
\vspace{-1.2em}

\section{Conclusion}
In this paper, we propose a QLook architecture for predicting the viewport in 6DoF environments. Our results demonstrate that the QLook framework achieves higher precision in viewport prediction. Showcasing the potential of quantum-classical hybrid models in advancing immersive VR experiences.
Identity block helps our solution in faster convergence. Our method is accurate and scalable for different users' data. To the best of our knowledge, QLook is the first work to study quantum applicability for VR-viewport prediction.
\nocite{*}

In the future, we will investigate a caching strategy based on QNN to further reduce latency in VR streaming settings. We also plan to extend our work to scenarios where multiple users interact simultaneously, moving beyond the current assumption of independent user behavior.
\vspace{-0.5em}
\bibliographystyle{IEEEtran}
\bibliography{main}

\end{document}